\documentclass[aps,prl,showpacs,twocolumn,floatfix,superscriptaddress]{revtex4}
\usepackage{graphicx}
\usepackage{amsmath}
\usepackage{amsfonts}

\begin{document}

\title{Self-propelled particles with fluctuating speed and direction of motion}

\author{Fernando Peruani} \email{peruani@pks.mpg.de} 
\affiliation{Max Planck Institute for the Physics of Complex Systems, N\"othnitzer Str. 38, 01187 Dresden, Germany}
\affiliation{ZIH, Technische Universit\"at Dresden, Zellescher Weg 12, 01069 Dresden, Germany}
\author{Luis G. Morelli}  
\affiliation{Max Planck Institute for the Physics of Complex Systems, N\"othnitzer Str. 38, 01187 Dresden, Germany}

\date{\today}

\begin{abstract}
We study general aspects of active motion with fluctuations in the speed and the direction of motion in two dimensions.
We consider the case in which fluctuations in the speed are not correlated to fluctuations in the direction of motion,  and assume that both processes can be described by independent characteristic time-scales. 
We show the occurrence of a complex transient that can exhibit a series of alternating regimes of motion, for two different angular dynamics which correspond to persistent and directed random walks.
We also show additive corrections to the diffusion coefficient.
The characteristic time-scales are also exposed in the velocity autocorrelation, which is a sum of exponential forms.   
%
%
\end{abstract}


\maketitle

The study of cell movement on surfaces can shed light on the processes that underly cell motility~\cite{bray}.
{\it In vitro} experiments that characterize cell movement include wound closure assays and individual cell tracking to determine cell trajectories~\cite{czirok_98,cameron_04,flyvbjerg_05,kumar_06}.
To interpret and understand such experiments it is necessary to rely on a phenomenological description of the motion, 
providing expressions that allow to fit the experimental observations and compute motility indexes~\cite{dickinson_93,flyvbjerg_05}.

Persistent motion subject to fluctuations has been described by a class of stochastic process known as persistent random walk~\cite{uhlenbeck_30,ebeling,schweitzer,othmer_88,mikhailov_97}.
In such processes the direction of motion fluctuates, but on short time-scales a persistence to move in the current direction is observed.
Formally, the velocity autocorrelation function $\langle \mathbf{v}(t) \cdot \mathbf{v}(0) \rangle$ exhibits a finite decay time,
giving rise to a ballistic regime $\langle \mathbf{x}^2(t) \rangle \sim t^2$ for short times and 
a crossover to a diffusive regime $\langle \mathbf{x}^2(t) \rangle \sim t$ for long times~\cite{uhlenbeck_30,othmer_88,mikhailov_97}.

A related problem is that of the directed motion of self-propelled particles driven by an external field.
Single cells can be directed by external signals in the form of molecule gradients,
as in the case of fibroblasts~\cite{schienbein_94} or the amoebae {\it D. discoideum}~\cite{schneider_06}.
Directed motion with fluctuations can be described by another broad class of stochastic process known as the directed random walk~\cite{othmer_88,schweitzer},
which displays a diffusive regime for short times followed by a ballistic regime for long times~\cite{huang_02, schienbein_94}.

In previous works either fluctuations in the speed and direction of motion were considered to occur simultaneously, 
or fluctuations of the speed were simply neglected~\cite{uhlenbeck_30,othmer_88,mikhailov_97,ebeling,schweitzer,tojo_96,bracher_04,komin_04}.
In this paper, we study two dimensional stochastic motion with uncorrelated fluctuations of the speed and the direction of motion,
assuming that both processes can be described by independent characteristic time-scales.
We investigate persistent and directed random walks, and derive exact expressions for the mean squared displacement and the asymptotic diffusion coefficient for arbitrary speed and angular stationary distributions.

\vspace{0.1cm} \noindent {\it General aspects.}
We consider self-propelled particles that move in two dimensions.
The velocity $\mathbf{v}(t)$ at time $t$ is represented by an angle $\theta(t)$, and a modulus ---the speed--- $v(t)$.
The dynamics of the velocity $\mathbf{v}(t)$ is given by a stochastic process which for the moment we do not specify.
We introduce $d(\theta, t)$, the total distance covered by the particle moving along the direction $\theta$ since the beginning of the trajectory at $t=0$.
Given a particular trajectory characterized by $\tilde{\theta}(t)$ and $\tilde{v}(t)$, we can express this distance as
$$d(\theta, t) =  \int_{0}^{t}{dt'} \int_{0}^{\infty} {dv \, v \,} {\delta(v - \tilde{v}(t')) \delta(\theta - \tilde{\theta}(t'))}.$$
The ensemble average of this distance is
\begin{equation} \label{eq:d}
\langle d(\theta,t) \rangle = \int_{0}^{t}{dt'} \int_{0}^{\infty} {dv \, v \,} { p(\theta,v,t')}  \, ,
\end{equation}
where we have introduced the probability density to find the particle moving in the direction $\theta$ with speed $v$ at time $t$,
$p(\theta,v,t) = \langle \delta(v - \tilde{v}(t)) \delta(\theta - \tilde{\theta}(t)) \rangle$~\cite{risken}.
Here we denote ensemble averages by $\langle \ldots \rangle$.
The correlations $\langle d(\theta',t) d(\theta'',t) \rangle$ between the distances
can be written in terms of the joint probability distribution.
We first express these correlations in terms of the ensemble average of delta distributions,
and use $p(\theta',v',t';\theta'',v'',t'') = \langle \delta(v - \tilde{v}(t')) \delta(v - \tilde{v}(t'')) \delta(\theta - \tilde{\theta}(t')) \delta(\theta - \tilde{\theta}(t'')) \rangle$
to obtain
\begin{eqnarray} \label{eq:dd}
\langle  d(\theta',t) d(\theta'',t) \rangle = \int_{0}^{t}{dt'}\int_{0}^{t}{dt''}   &    \\
                         \int_{0}^{\infty}{dv'} \int_{0}^{\infty}{dv''}{v' v''}   &   {p(\theta',v',t';\theta'',v'',t'')} \, . \nonumber
\end{eqnarray}

We can use the distance $d(\theta,t)$ to express the position of the particle at time $t$ as
$\mathbf{x}(t) =  \int_{-\pi}^{\pi} {d(\theta, t) \check{r}(\theta) d\theta }$,
where $\check{r}(\theta) = \cos(\theta)\check{x} + \sin(\theta) \check{y}$ is the unit vector along the direction $\theta$.
Then the mean value of the position is
\begin{equation} \label{eq:x}
\langle \mathbf{x}(t) \rangle = \int_{-\pi}^{\pi} \langle d(\theta,t) \rangle \check{r}(\theta) d\theta \, ,
\end{equation}
and the mean square displacement
\begin{equation} \label{eq:xx}
\langle \mathbf{x}^2(t) \rangle =  \iint_{-\pi}^{\pi}{d\theta'}{d\theta''} \langle d(\theta',t) d(\theta'',t) \rangle \check{r}(\theta') \cdot \check{r}(\theta'').
\end{equation}

Eqs.~(\ref{eq:d}) to (\ref{eq:xx}) provide a general way to
calculate the mean value of the position and the mean square
displacement, which so far does not involve any assumptions. In the
following, we consider the special case in which the fluctuations in
the speed $v$ are not correlated with the fluctuations in the
direction $\theta$. As a consequence, $p(\theta,v,t) = p(\theta,t)
p(v,t)$ and $p(\theta',v',t';\theta'',v'',t'') = p(\theta',t';\theta'',t'') p(v',t';v'',t'')$. 
Such a situation could naturally arise if fluctuations of the speed are endogenous and produced by an irregular engine,
while fluctuations of the direction of motion are produced by random changes in the environment.
In the following, we further assume that the speed fluctuations are in the stationary state with an arbitrary speed distribution $p(v,t) = \rho(v)$,  and the joint probability is given by 
\begin{eqnarray} \label{eq:assumption}
p(v',t' ; v'',t'') &=& \rho(v'') \delta(v'-v'') e^{-\beta |t'-t''|} \\
                   &+& \rho(v'')  \rho(v')  \left(1-e^{-\beta|t'-t''|}\right) \, .  \nonumber
\end{eqnarray}
This expression for the joint probability distribution implies that the speed correlations decay exponentially as $\langle v(t) v(0) \rangle - \langle v \rangle^2 = (\langle v^2 \rangle - \langle v \rangle^2) e^{-\beta t}$.
Eq.~(\ref{eq:assumption}) describes particles that keep on moving with roughly the same speed for a characteristic time $\beta^{-1}$, while for larger times the values of the speed become uncorrelated.
An example of a stochastic process which generates such statistics is given by $v(t) = \widetilde{\eta}(t)$, 
where the value of $\widetilde{\eta}(t)$ is taken from a distribution $\rho (v)$ with waiting times given by a Poisson process of rate $\beta$. 
The evolution Eq. for the probability density $p(v,t)$  can be expressed as $\partial_{t}p(v,t)=-\beta p(v,t) + \beta \rho(v)$, leading to the conditional probability given by Eq. (\ref{eq:assumption}).

Under these assumptions Eqs.~(\ref{eq:d}) and (\ref{eq:dd}) can now be simplified performing the integrals on the speed:
\begin{equation} \label{eq:dv}
\langle d(\theta,t) \rangle = \langle v \rangle \int_{0}^{t} {dt' p(\theta,t')} \, ,
\end{equation}
\begin{eqnarray}  \label{eq:ddv}
\langle d(\theta',t) d(\theta'',t) \rangle = \langle v \rangle^2  \iint_{0}^{t} {dt' dt'' p(\theta',t';\theta'',t'')} \\ \nonumber
+ (\langle v^2 \rangle - \langle v \rangle^2 ) \iint_{0}^{t} {dt' dt'' p(\theta',t';\theta'',t'')} e^{-\beta |t'-t''|} \, .
\end{eqnarray}

\vspace{0.1cm} \noindent {\it Persistent random walk.} 
As a first application, we consider the case of a persistent random walk. 
We study a problem in which the angular probability distribution function obeys a diffusion equation characterized by the diffusion constant $\kappa$.
An example of a stochastic process described by such an equation is given by $\dot{\theta}(t) = \eta(t)$, where $\eta(t)$ is an uncorrelated white noise. 
We assume that particles start moving from the origin in all possible directions with equal probability, so $p(\theta,t)=1/2\pi$ for all times and $\langle \mathbf{x}(t) \rangle = \mathbf{0}$.
However a given particle starts moving along a particular direction and smoothly explores other directions, 
so a characteristic time must elapse before we can find this particle pointing with equal probability in any direction. 
This is described by the conditional probability distribution $p(\theta',t'|\theta'',t'')$, which obeys the diffusion equation
$\partial_{t'}p(\theta',t'|\theta'',t'')=\kappa\partial_{\theta'\theta'}p(\theta',t'|\theta'',t'')$, 
with the initial condition $p(\theta',t'|\theta'',t') = \delta(\theta'-\theta'')$.
To warrant the conservation of the probability we impose the periodic boundary condition 
$p(\pi,t'|\theta'',t'') = p(-\pi,t'|\theta'',t'')$ and $\partial_{\theta' } p(\pi,t'|\theta'',t'') = \partial_{\theta'}p(-\pi,t'|\theta'',t'')$.
The solution for the conditional probability is 
\begin{equation} \label{eq:pq.drw}
p(\theta',t' | \theta'',t'') = \frac{1}{2 \pi} + \frac{1}{    \pi} \sum_{m=1}^{\infty}{ \cos \left[m (\theta'-\theta'')\right] \, e^{-m^2 \kappa|t'-t''|}} \, .
\end{equation}
As $|t'-t''| \to \infty$ the information about the direction of motion at time $t''$ is completely lost, and the conditional probability approaches the asymptotic value $1/2\pi$.
The slowest mode $m=1$ sets the characteristic time-scale $\kappa^{-1}$ that describes the duration of the transient. 

\begin{figure}[t]
\centering\resizebox{7.5cm}{!}{\rotatebox{0}{\includegraphics{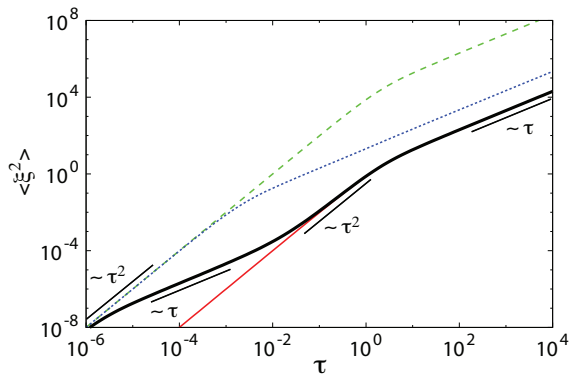}}}
\caption{Scaled mean squared position as a function of scaled time for
persistent random walks. The lines correspond to Eq.~(\ref{eq:xx.prw}) for rescaled variables.
The thin solid red line corresponds to $\mu=0$, and $\mu=100$ for the other curves, 
with $\gamma=10^{-3}$ (dashed green line), 
$\gamma=10^3$ (dotted blue line) and 
$\gamma=10^6$ (thick solid black line).}
\label{fig:prw.xx}
\end{figure}

Recalling that $p(\theta',t';\theta'',t'')=p(\theta',t'|\theta'',t'')p(\theta'',t'')$ and using Eq.~(\ref{eq:pq.drw}) in Eq.~(\ref{eq:ddv}), 
we obtain from Eq.~(\ref{eq:xx}) that
\begin{eqnarray} \label{eq:xx.prw}
\langle \mathbf{x}^2(t) \rangle &=& 2 \frac{\langle v \rangle ^2}{\kappa^2} \left( \kappa t -1+e^{-\kappa t} \right) \\
&+& 2 \frac{\langle v^2 \rangle - \langle v \rangle ^2}{(\kappa+\beta)^2} \left( (\kappa+\beta) t -1+e^{-(\kappa+\beta) t} \right)  \, . \nonumber
\end{eqnarray}
In the absence of speed fluctuations, the speed variance $\sigma^2 = \langle v^2 \rangle - \langle v \rangle ^2$  vanishes
and Eq.~(\ref{eq:xx.prw}) reduces to the well known result for persistent Brownian particles~\cite{uhlenbeck_30},
which exhibits a single crossover at $t \sim \kappa^{-1}$, see thin solid red line in Fig.~\ref{fig:prw.xx}.
When fluctuations become relevant, a previous crossover from a quadratic to a linear regime occurs at $t\sim~(\kappa+\beta)^{-1}$.
If the separation of time-scales allows it another crossover can be observed  between these two,
when the linear regime of the second term turns into the quadratic regime of the first one, see the thick solid black line in Fig.~\ref{fig:prw.xx}.

In order to unveil the different regimes that Eq.~(\ref{eq:xx.prw}) permits,
we introduce non-dimensional variables $\xi = x \kappa / \langle v \rangle$ and $\tau = \kappa t$, and parameters
$\mu=\sigma / \langle v \rangle$ and $\gamma = \beta / \kappa$.
For $\gamma \ll 1$ there is a single crossover at $\tau \sim 1$, see dashed green line in Fig.~\ref{fig:prw.xx}.
For larger values of $\gamma$, solutions lie between the dashed green line and thin solid red line.
For $\tau \ll (1+\gamma)^{-1}$ we observe $\langle \xi^2 \rangle \simeq (1+\mu^2)\tau^2$.
A first crossover occurs at $\tau_1 \sim  (1+\gamma)^{-1}$.
For larger times, if the separation of time-scales is such that $(1+\gamma)^{-1} \ll \tau \ll 1$ then $\langle \xi^2 \rangle \simeq \tau^2 + 2 \mu^2 \tau / (1+\gamma)$.
Provided that $\mu^2$ is sufficiently large, a second crossover occurs at $\tau_2 \sim 2 \mu^2 / (1+\gamma)$
separating a transient linear regime from a second quadratic regime.
Finally, for $\tau \gg 1$ the asymptotic diffusive regime emerges with $\langle \xi^2 \rangle \simeq 2 (1+\mu^2/(1+\gamma)) \tau$,
after the third crossover at $\tau_3 \sim 1$. 
Such asymptotic regime can be described in terms of an effective diffusion coefficient, defined as ${\cal D}=\lim_{t \to \infty} ( \langle \mathbf{x}^2(t)\rangle - \langle \mathbf{x}(t) \rangle ^2 ) / 2t$.
From Eq. (\ref{eq:xx.prw}) we obtain
${\cal D} ={\langle v \rangle ^2}{\kappa^{-1}} + {\left( \langle v^2 \rangle - \langle v \rangle ^2 \right)}{(\kappa + \beta)^{-1}} \,.$
Speed fluctuations introduce an additive correction to the well known diffusion coefficient for constant speed~\cite{mikhailov_97}, 
and can lead up to four consecutive regimes of motion separated by three crossovers, see Fig.~\ref{fig:prw.xx}. 
In the absence of speed fluctuations only one crossover is found~\cite{othmer_88,mikhailov_97}.


In Fig.~\ref{fig:prw.xx} we use a large speed variance and plot $\langle \mathbf{x}^2(t) \rangle$ over a huge range to illustrate all the features of Eq.~(\ref{eq:xx.prw}).
Below we discuss experimental constraints in the observation of the phenomena described here.

\vspace{0.1cm} \noindent {\it Directed persistent random walk.}
As a second example we consider the directed random walk, in which the particles have some preferred direction of motion.
This could be the case for particles moving in a symmetry-breaking field or gradient. 
We assume that angular fluctuations are in the stationary state $p(\theta,t)=\rho(\theta)$.
We describe the presence of an external field by assuming that $\int_{-\pi/2}^{\pi/2} {\rho(\theta) d\theta} > 1/2$ 
together with the symmetry requirement $\rho(-\theta) = \rho(\theta)$,
setting a preferred direction of motion along $\theta=0$.
Time correlations decay exponentially with a characteristic time $\alpha^{-1}$
\begin{eqnarray}
p(\theta',t' ; \theta'',t'') &=& \rho(\theta'') \delta(\theta'-\theta'') e^{-\alpha |t'-t''|} \\ \nonumber
                                     &+& \rho(\theta'') \rho(\theta') \left(1-e^{-\alpha|t'-t''|} \right) \, .
\end{eqnarray}
A realization of such stochastic process is $\theta(t) = \eta(t)$, 
where the value of the noise $\eta(t)$ is taken from a distribution $\rho (\theta)$ at times given by a Poisson process of rate $\alpha$. 
Using this expressions for the angular probability distributions together with Eqs.~(\ref{eq:dv}) and (\ref{eq:ddv}) in Eqs.~(\ref{eq:x}) and (\ref{eq:xx})
we arrive at the following expressions for the mean value of the position, $\langle \mathbf{x}(t) \rangle = \sqrt{c} \langle v \rangle  t \, \check{x}$,
and the mean square displacement
\begin{eqnarray} \label{eq:xx.drw}
\langle \mathbf{x}^2(t) \rangle &=& \langle v \rangle ^2 \left[ ct^2 + 2 (1-c) \varphi_{\alpha}(t) \right] \\
&+& 2 \left(  \langle v^2 \rangle - \langle v \rangle ^2 \right) \left[ (1-c) \varphi_{\alpha+\beta}(t) + c \varphi_{\beta}(t)  \right]   \nonumber
\end{eqnarray}
where $c = \langle \cos \theta \rangle^2$ and $\varphi_{\alpha} (t) = \alpha^{-2} \left[ \alpha t - \left( 1 - e^{-\alpha t} \right) \right]$.
\begin{figure}[t]
\centering\resizebox{7.5cm}{!}{\rotatebox{0}{\includegraphics{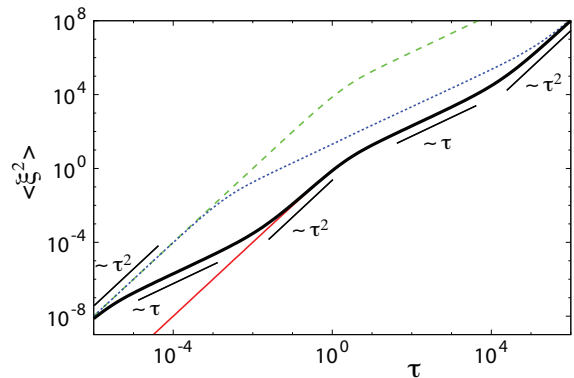}}}
\caption{Scaled mean squared position as a function of scaled time for directed random walks. 
The lines correspond to Eq.~(\ref{eq:xx.drw}) for the rescaled variables, with $c=10^{-4}$. The thin solid red line corresponds to $\mu = 0$. 
The other curves correspond to $\mu=100$,
with $\gamma=10^{-3}$ (dashed green line), 
$\gamma=10^3$ (dotted blue line) and 
$\gamma=10^6$ (thick solid black line).} 
\label{fig:brw.xx}
\end{figure}
%
The effective diffusion coefficient ${\cal D}$ as defined above also exhibits an additive correction, 
${\cal D} = (1-c) \langle v \rangle ^2  \alpha^{-1} +( \langle v^2 \rangle - \langle v \rangle ^2)[(1-c)( \alpha + \beta)^{-1} + c \beta^{-1}] .$
Speed fluctuations introduce a new time-scale which together with the independent time-scale of angular fluctuations can lead up to five alternating regimes of motion separated by four crossovers, see Fig.~\ref{fig:brw.xx}. 
In the absence of speed fluctuations the second line in Eq.~(\ref{eq:xx.drw}) vanishes and the mean square displacement exhibits only two regimes~\cite{huang_02}.

\begin{figure}[t]
\centering\resizebox{\columnwidth}{!}{\rotatebox{0}{\includegraphics{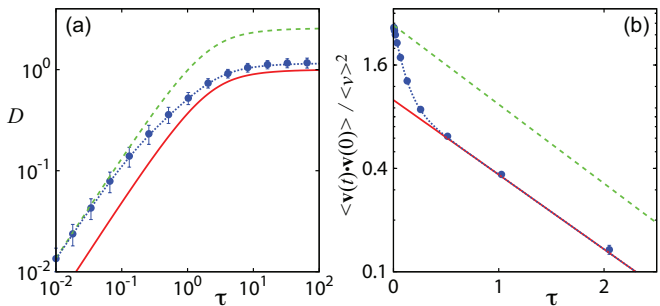}}}
\caption{(a) Re-scaled mean squared displacement and (b) velocity autocorrelation as a function of scaled time for persistent random walks. 
The lines in (b) correspond to Eq.~(\ref{eq:vv}) for rescaled variables.
In both panels, the solid red line corresponds to $\mu=0$. For the other curves $\mu=1.31$, 
with $\gamma=10^{-1}$ (dashed green line) and $\gamma=10$ (dotted blue line).
Dots correspond to numerical simulations as described in the text.} 
\label{fig:exp}
\end{figure}

\vspace{0.1cm} \noindent {\it Concluding remarks.} 
The interplay of speed and angular fluctuations gives rise to a sequence of regimes, revealing a complex transient not observed when speed fluctuations are absent.
The occurrence of such a complex transient can wreck the interpretation of experimental observations, due to the constrains imposed by resolution and finite size limitations. 
Particle size ---or a fluctuating cell shape--- sets the smallest accessible length-scale, while the field of the experimental setup sets the largest.
The temporal window is similarly bounded.
%
If the window of observation is limited to a part of the complex transient, anomalous diffusion could be wrongly interpreted.
Superdiffusion has been repeatedly reported from experimental data~\cite{viswanathan96, libchaber99}. 
However, distinguishing true superdiffusion from a persistent or directed random walk is a subtle task~\cite{viswanathan05}.
Our results suggest that in some cases the observed anomalous behavior could be related to one or more of the reported crossovers.
%
In Fig.~\ref{fig:exp} (a) we display $D(\tau) =  ( \langle \mathbf{\xi}^2(\tau)\rangle - \langle \mathbf{\xi}(\tau) \rangle ^2 ) / 2 \tau$
for a persistent random walk, using time and space ranges which are reasonable for current experimental setups~\cite{czirok_98,flyvbjerg_05}.
Furthermore, we choose $\mu=1.31$ according to data reported in~\cite{flyvbjerg_05}, and values of $\gamma$ within experimental ranges. 
The solid red line is the result without speed fluctuations.
Dots correspond to numerical simulations performed with $\mu=1.31$ and $\gamma = 10$.
In the simulations speeds are chosen at a rate $\beta = 4.0$ h$^{-1}$ from a speed distribution $\rho(v)\sim v^{-3/2}$ for $v\in[1,v_c]$ and zero otherwise, with $v_c$ such that $\mu = 1.31$. 
Angles are chosen at a rate $9.6$ h$^{-1}$ from a uniform distribution of width $1$ rad centered around the direction of motion,
and so yielding $\kappa = 0.4$ h$^{-1}$.
Error bars are the standard deviation from the mean value obtained for 100 realizations with 100 particles each.
This means that a particular 100 particles experiment should fall within the range of such error bars.

The two characteristic time-scales of the system are also exposed in the velocity autocorrelation function, which is given by a sum of two exponentials
\begin{equation} \label{eq:vv}
\langle \mathbf{v}(t) \cdot \mathbf{v}(0) \rangle =  \langle v \rangle^2 \, e^{-\kappa t} +  (\langle v^2 \rangle - \langle v \rangle^2) \, e^{-(\beta+\kappa) t} \, .
\end{equation}
Autocorrelations of similar functional form have been observed in cell motility experiments, but the microscopic origin of the two time-scales has not been established~\cite{flyvbjerg_05}.
Here we show that independent fluctuations in speed and direction could produce such autocorrelations, see Fig.~\ref{fig:exp}~(b).
Fast intracellular processes could give rise to such fluctuations in speed with small characteristic time-scales~\cite{joanny03,ponti04}.
The simulations suggest that fluctuations in speed as the ones observed in experiments might be enough to cause visible deviations from the classical result~\cite{uhlenbeck_30}.
In the case of directed motion, the presence of an external field decouples the time-scale of speed fluctuations and the velocity autocorrelation results in a sum of three exponentials:
$\langle \mathbf{v}(t) \cdot \mathbf{v}(0) \rangle = (\langle v \rangle^2 + \sigma^2 e^{-\beta t}) [c+(1-c) e^{-\alpha t}] $.
In this case we are not aware of experiments showing such autocorrelations.
While here we have considered the case in which speed and angular fluctuations are not correlated,
the case in which they are is also of much interest and deserves future attention.

We thank S. F. N\o{}rrelykke and B. Lindner for insightful comments and valuable suggestions on the manuscript.
FP thanks M. B\"ar and A. Deutsch for support and acknowledges funding from the Deutsche Forschungsgemeinschaft through Grant No. DE842/2.

\end{document}